\documentclass[12pt]{article}
\usepackage{amsmath}
\numberwithin{equation}{section}

\begin{document}
\newcommand{\ltwid}{\mathrel{\raise.3ex\hbox{$<$\kern-.75em\lower1ex\hbox{$\sim$}}}}
\newcommand{\gtwid}{\mathrel{\raise.3ex\hbox{$>$\kern-.75em\lower1ex\hbox{$\sim$}}}}

\title{What Connects Different Interpretations\\
 of Quantum
Mechanics?\footnote{\rm To appear in the proceedings of the workshop {\sl
Quo Vadis Quantum Mechanics}, Center for Frontier Sciences, Temple
University, Philadelphia, PA, September 24--27, 2002} }

\author{James B.~Hartle\thanks{hartle@physics.ucsb.edu}\\
Department of Physics, University of California\\ 
Santa Barbara, California 93106-9530}

\date{}

\maketitle

\begin{abstract}

We investigate the idea that different interpretations of quantum mechanics
can be seen as restrictions of the consistent (or decoherent) histories
quantum mechanics of closed systems to particular classes of histories,
together with the approximations and descriptions of these histories that
the restrictions permit.

\end{abstract}



\setlength{\baselineskip}{.2in}

\section{Introduction}

A list of some thirteen
different interpretations of quantum mechanics
was discussed at the concluding session of this conference\footnote{See the 
contribution of A.~Elitzur to this volume.} . The very length of
this list invites the questions: What are
the relationships between these interpretations?; To what uses may they
be put?; and Is it possible to objectively settle on one?. This brief
article offers some personal reflections on these questions.

The defining thread connecting interpretations of quantum theory is their
agreement on the probabilities for the outcomes of measurements, at least to
an excellent approximation.  Some
formulations may provide probabilities for further kinds of alternatives such
as the position of the Moon when it is not receiving attention from
observers, or the values of density fluctuations in the very early
universe when there were no observers around.  However, a formulation 
that does not reproduce the standard
textbook answers for the probabilities of {\it measurements} it is not an
interpretation of {\it quantum mechanics}.  Rather, it is a different
theory.  Such alternatives to quantum theory are of great
interest but not the subject of this essay.

The idea explored here is that a number of different
interpretations of quantum mechanics can be connected through the
consistent (or decoherent) quantum mechanics of a closed system. 
Specifically, a number of interpretations can be
seen as {\it restrictions} of consistent histories quantum theory to
particular kinds of sets of alternative histories\footnote{``Frameworks''
in the terminology of Griffiths.} together with the approximations and
special descriptions of the sets that these restrictions permit.
This essay examines three cases where this connection can be made and gives
brief discussions of the utility of the restrictions involved.

\section{The Quantum Mechanics of Closed Systems}

We begin with a very brief review of the quantum mechanics of a closed
system, most generally the universe as a whole\footnote{For details see
the classic expositions in \cite{Gri02,Omn94,Gel94}.}. 
To simplify the discussion
we neglect quantum gravity and assume a fixed background spacetime
geometry. The familiar apparatus of Hilbert space, states, and operators
may then be employed to formulate the quantum mechanics of the closed
system\footnote{For the generalizations that may be otherwise required, see
\cite{Har95c}. We view the quantum mechanics of closed systems
as an extension and completion of
the Everett formulation and therefore do not count that as a separate
interpretation.}. As a simple model, we can think of a large, isolated box
of $N$ non-relativistic particles. Dynamics can be specified in terms of
particle positions $\vec x_i$ and momenta $\vec p_i$ by a Hamiltonian
\begin{equation}
H= \sum\limits^N_{i=1}\ \frac{\vec p^2_i}{2m_i} + V (\vec x_i).
\label{twoonea}
\end{equation}
Both observers and observed, if any, are contained inside.  This is
evidently not the most general description of a closed system but it will
suffice to illustrate some of the connections between interpretations that
we describe later.
 
We take the closed system to be described by a quantum state 
$|\Psi\rangle$. The most general objective of quantum theory is the
prediction from $H$ and $|\Psi\rangle$ of the probabilities of the 
individual members of a set of
coarse-grained alternative histories of the closed system.  A history is
described by giving a sequence of alternatives $(\alpha_1, \cdots
\alpha_n)$ at a series of times $t_1, \cdots, t_n$. Alternatives at a
moment of time $t_k$ are represented by an exhaustive set of orthogonal,
Heisenberg picture, projection operators $\{P^k_{\alpha_k} (t_k)\}$ and 
a history of alternatives is represented 
by the corresponding chain of projections called a {\it class
operator}
\begin{equation}
C_\alpha = P^k_{\alpha_n} (t_n) \cdots P^1_{\alpha_1} (t_1).
\label{twoone}
\end{equation}
(On the left side of \eqref{twoone}
we have abbreviated the whole chain $(\alpha_1, \cdots,
\alpha_n)$ by a single index $\alpha$). For example, if we are interested in
a history of the Earth moving around the Sun, the $P$'s might be
projections onto exclusive ranges of the center of mass position of the
Earth at a sequence of times.  This set of histories is {\it
coarse-grained}\ because alternatives are not specified at every time but
only at some times, because the center of mass position is not specified
exactly but only in certain ranges, and because not every variable
describing  the universe is specified but only the center of mass of the
Earth.

The class operators $C_\alpha$ defined in (\ref{twoone}) permit the 
construction of {\it branch state vectors}
\begin{equation}
|\Psi_\alpha\rangle = C_\alpha |\Psi\rangle
\label{twotwo}
\end{equation}
for each history in the coarse-grained set.  A set of histories {\it
decoheres} when there is negligible mutual interference between all the
branch state vectors:
\begin{equation}
\langle\Psi_\alpha |\Psi_{\alpha^\prime}\rangle \approx 0, \ \alpha
\not= \alpha^\prime.
\label{twothree}
\end{equation}
The joint {\it probability} $p_\alpha$ for all the events in a history $\alpha$
is
\begin{equation}
p_\alpha = \Vert\ |\Psi_\alpha\rangle\ \Vert^2 = \Vert C_\alpha |\Psi\rangle
\Vert^2
\label{twofour}
\end{equation}
when the set of histories decoheres. Decoherence ensures the validity of
the probability sum rules which are among the defining properties of
probability.

The above discussion is brief, certainly oversimplified in some respects,
but sufficient we hope for understanding the remarks which follow.  The key
point for the ensuing discussion is the following: Decoherent histories
quantum mechanics predicts probabilities for many different sets of
alternative histories which are {\it complimentary} in the following sense:
Each set is part of a complete quantum description of the system, but there is
no fine-grained decoherent set histories of which all the decoherent sets
are coarse-grainings. A set of histories coarse-grained by the Earth's
center of mass momentum is an example of a set which (if decoherent) would
be complementary to the set coarse-grained by the Earth's center of mass
position.  

Given $H$ and $|\Psi\rangle$, it is in principle
possible to calculate all decoherent sets. Among these is the
quasiclassical realm of everyday experience, coarse-grained by the variables
of classical physics, and exhibiting classical patterns correlation in time
summarized approximately by classical equations of motion.  As human observers 
we focus
almost entirely on coarse-grainings of this quasiclassical realm.  However,
quantum theory does not distinguish the quasiclassical realm from other
decoherent sets except by properties such as its classicality.

The picture of quantum reality which emerges from the quantum mechanics of
closed systems is very different from the reality of classical physics
involving, as it does, many {\it complementary} descriptions of the
universe that are mutually incompatible.  Restricting the allowed sets of
histories by some principle\footnote{A {\it set selection principle} in
terminology of Dowker and Kent \cite{DK96}.} typically yields a description
of reality that is closer in character to the familiar classical one.  
We will see that
in the cases to be discussed. 

\section{Three Case Studies}

This section considers the idea offered in the Introduction for three
different interpretations of quantum theory.

\subsection{Copenhagen Quantum Mechanics}

The Copenhagen quantum mechanics found in most textbooks is concerned with
the probabilities of histories of the outcomes of {\it measurements} carried 
out by {\it observers}.  The subsystem being observed is described by a 
Hilbert space
${\cal H}_s$. Dynamics is specified by a Hamiltonian $h$ acting on ${\cal
H}_s$ when the subsystem
is isolated.  Initially the subsystem is assumed to be a state
$|\psi\rangle$ in ${\cal H}_s$.  The outcomes of a measurement carried out
at time $t_k$ can be described by a set of orthogonal, Heisenberg picture,
projection operators $\{s^k_{\alpha_k} (t_k)\}$, $\alpha_k=1, 2, \cdots$
analogous to the $P$'s described in Section II.  The probabilities for a
history of ideal measurements (ones that disturb the subsystem as little as
possible) at times $t_1, \cdots, t_n$ are given by the analog of
(\ref{twofour}).
\begin{equation}
p_\alpha = \left\Vert s^n_{\alpha_n} (t_n) \cdots s^1_{\alpha_1} (t_1)
|\psi\rangle \right\Vert^2.
\label{threeone}
\end{equation}
Consistency is not an issue for these probabilities.  Probabilities for a
coarser-grained history need not be the sum of the probabilities of
finer-grained histories
consistent with it.  Finer and coarser grained measurements correspond to
different interactions of the subsystem with an external apparatus.  Sets
of histories describing alternative measurements do not have to decohere. 

Copenhagen quantum mechanics is an approximation to the quantum mechanics
of closed systems that is appropriate for histories of measurement
situations when the decoherence of alternatives that
register the outcomes of the measurements can be idealized as exact.  We
sketch only the essential features of a demonstration which are essentially
the same as many measurement models.  For details see,
{\it e.g.}~\cite{Har91a}, Section II.10.

We consider a closed system with a Hilbert space 
${\cal H}_s \otimes {\cal H}_r$
where ${\cal H}_s$ is the Hilbert space of the measured subsystem and the
${\cal H}_r$ is the Hilbert space of the rest of the universe including any
measuring apparatus and observers.  We assume an initial state of the form
$|\Psi \rangle = |\psi \rangle \otimes |\Phi_r\rangle$ and consider a
sequence of measurements at a series of times $t_1, \cdots, t_n$. Measured 
alternatives 
of the subsystem are described by projection operators whose Schr\"odinger 
picture representatives have the form $S^k_{\alpha_k} = s^k_{\alpha_k} \otimes
I_r$. In a typical measurement situation, an alternative such as
$S^k_{\alpha_k}$ becomes correlated with an alternative of the apparatus
and in particular with persistent records of the measurements.  The
orthogonality and persistence of these records guarantees the decoherence
of the histories of measured outcomes.  If the usual assumption is made
that the measurement interaction disturbs the subsystem as little as
possible (ideal measurement), then
\begin{equation}
p_\alpha = \left\Vert S^n_{\alpha_n} (t) \cdots S^1_{\alpha_1} (t_1)
|\Psi\rangle \right\Vert^2_{\cal H} \approx \left\Vert s^n_{\alpha_n} (t_n)
\cdots s^1_{\alpha_1} (t_1) |\psi\rangle\right\Vert^2_{{\cal H}_r}.
\label{threetwo}
\end{equation}
Thus Copenhagen quantum mechanics is recovered as a restriction of, and
approximation to, the quantum mechanics of closed systems.  The second
equality in (\ref{threetwo}) is not exact but true to an excellent
approximation in realistic measurement situations --- typically far beyond the
accuracy which the probabilities can be checked or the physical situation
modeled. 

The utility of the approximate quantum mechanics of measured
subsystems is evident.  It is a truly excellent approximation for every
laboratory experiment which has tested the principles of quantum theory.
Further, the calculations of the approximate Copenhagen probabilities
utilizing just the Hilbert space of the measured subsystem will generally
be vastly simpler than in the Hilbert space of the universe. These
advantages, however, should not obscure the utility of embedding the
Copenhagen quantum mechanics in the more general quantum mechanics of
closed systems for understanding measurements (as above) and calculating just
how good an approximation it is.

\subsection{Bohm Theory}

To summarize the features of Bohm theory \cite{BH93} that are relevant to
the present discussion, it is convenient to restrict attention to the closed
system consisting of $N$, non-relativistic particles in a box discussed in
Section II.
An initial wave function $\Psi(\vec x_1, \cdots \vec x_N, 0)$ is given.
The particles in the box move on trajectories $\vec x_i (t)$
that obey two deterministic
equations.  The first is the Schr\"odinger equation for $\Psi$:
\begin{equation}
i\hbar\ \frac{\partial \Psi}{\partial t} = H\Psi.
\label{threesix}
\end{equation}
Then, writing $\Psi = R\ \exp(iS)$ with $R$ and $S$ real, the second
equation is the deterministic equation for the $\vec x_i (t)$
\begin{equation}
m_i\ \frac{d\vec x_i}{dt} = \vec \nabla_{\vec x_i} S\, (\vec x_1, \cdots,
\vec x_N).
\label{threeseven}
\end{equation}
The initial wave function is the initial condition for (\ref{threesix}).
The theory becomes a statistical theory with the assumption that the {\it
initial values} of the $\vec x_i$ are distributed according to the
probability density on configuration space
\begin{equation}
\wp \left(\vec x_1, \cdots, \vec x_N\, ,\, 0\right) = \left|\Psi \left(\vec
x_1, \cdots, \vec x_N\, ,\, 0\right)\right|^2, \ \textrm{at the initial
time}\ 0.
\label{threeeight}
\end{equation}
Once this initial probability distribution is fixed, the probability of any
later alternatives is fixed by the deterministic equation
(\ref{threeseven}).

A coarse-grained Bohmian history $\alpha \equiv (\alpha_n, \cdots, \alpha_1)$ is
defined by a sequence of ranges $\{\Delta^k_{\alpha_k}\}$ of the $\vec x_i$
at a series of times $t_1, \cdots t_n$ and consists of the set of Bohmian 
trajectories $\vec x_i(t)$ that cross those ranges at the specified times.

The predictions of Bohm theory and the quantum mechanics of closed systems
can be compared for sets of alternative histories coarse-grained by ranges
of the position $\vec x_i$ at different times as above.  Generally
different
probabilities are predicted for the same set of histories \cite{Har02}. 
This difference
arises as follows: Bohm histories are deterministic.  That means that the
probability that the particles traverse a series of regions of
configuration space at a sequence of times is the same as the probability
of the initial values of $\vec x_i$ that evolve to those trajectories under
the equations of motion (\ref{threesix}) and (\ref{threeseven}).  
The probability of a Bohm
trajectory can therefore be represented as
\begin{equation}
p^{(\rm BM)}_\alpha = \left\Vert B_\alpha |\Psi\rangle \right\Vert^2
\label{threenine}
\end{equation}
where $B_\alpha$ is a {\it projection} onto the appropriate initial
conditions.

The probabilities of the {\it same} set of histories would be calculated
in decoherent histories quantum theory 
from [\textit{cf.}~(\ref{twofour})]
\begin{equation}
p^{(\rm DH)}_\alpha = \Vert C_\alpha |\Psi\rangle \Vert^2
\label{threeten}
\end{equation}
provided the set
is decoherent. Here the $C_\alpha$ are chains of projections like
(\ref{twoone})  
It is a simple observation is that a chain of projections like
(\ref{twoone}) is not generally a projection and that therefore $p^{(\rm BM)}_\alpha$
will not agree generally with $p^{\rm (DH)}_\alpha$. (See \cite{Har02} for
examples and further discussion.)

Another way of seeing the difference is to note that in Bohm theory the
wave function always evolves by the Schr\"odinger equation --- unitary 
evolution.
But the action of a chain of projections $C_\alpha$ on the initial state
can be described as unitary evolution interrupted by the action of the
projections (reduction).

Only in the case of histories with alternatives at a single
time are the predictions of Bohm theory and the quantum mechanics of closed
systems guaranteed to agree.  Then the $C_\alpha$ {\it are} projections.
But this is an important case because it leads to the conclusion that Bohm
theory and the quantum mechanics of closed systems agree on the
probabilities of the outcome of {\it measurements}.

One characteristic of a measurement situation which seems generally agreed
upon is that the results of a measurement are {\it recorded} --- at least
for a time.  A history $C_\alpha$ of measurement outcomes is {\it recorded}
in a set of alternatives $\{R_\alpha\}$ at a time later than the last
alternative in $C_\alpha$ if the values of the $R_\alpha$ are correlated with
the outcomes of the measurements described by the $C_\alpha$. The $R_\alpha$'s
are projections even if the $C_\alpha$'s are not. Bohm theory and the
quantum theory of closed systems will therefore agree on the probabilities
of these records.

Bohm theory can therefore be seen as a restriction of the quantum theory of
closed systems to alternatives describing the records of measurements (in
the $\vec x$'s) together with the description of these outcomes in
terms of deterministic trajectories obeying (\ref{threesix}) and
(\ref{threeseven}).  An
advantage of Bohm theory (that is, of this restriction) that we believe 
would be claimed by its proponents is the
clear specification of one set of histories (of the $\vec x$'s) as
preferred over others.  A potential disadvantage is that these histories,
although deterministic, may not be classical even in situations where the
correlations of classical physics in time are predicted with high
probability by the quantum mechanics of closed systems \cite{Gri99}.  
Thus, for example,
even when a classical past is retrodicted from present records from the
quantum mechanics of closed systems, Bohm theory may predict a
non-classical one depending on the nature of the initial condition
\cite{Har02}. 
 
\subsection{Sum-Over-Histories}

The starting point for a sum-over-histories formulation of quantum
mechanics is the specification of one set of fine-grained histories.  For
the model universe of non-relativistic particles in a box, these are the
particle paths $\vec x_i (t), i=1, \cdots, N$.  The allowed
coarse-grainings are partitions of this set of fine-grained histories into
an exhaustive set of exclusive classes.  For example, the paths could be
partitioned by how they traverse a set of regions of configuration space
$\{\Delta_{\alpha_k}\}$, $\alpha_k=1, 2, \cdots$ at sequence of times
$t_k$, $k=1, \cdots, n$.  The class operators $C_\alpha$ are specified by
giving their matrix elements as sums over the fine-grained paths in the
coarse-grained class labeled by $\alpha$. Denoting a point in the
$3N$-dimensional configuration space by $x$, this sum is
\begin{equation}
\left\langle x^{\prime\prime}|C_\alpha|x^\prime\right\rangle = \int_\alpha
\delta x\, e^{iS[x(t)]/\hbar}.
\label{threethree}
\end{equation}
Here, $S[x(t)]$ is the action functional and the sum is over all
fine-grained histories in the class labeled by $\alpha$.  For instance, in
the partition by sequences of sets of regions at a series of times, a
coarse-grained history $\alpha$ is labeled by the regions $(\alpha_1,
\cdots, \alpha_n)$ crossed at the sequence of times and the sum in
(\ref{threethree}) defining the class operator is over paths that cross 
these regions.  The
construction of probabilities is then as described in Section II.

Sum-over-histories quantum theory is evidently a restriction of the quantum
mechanics of closed systems described in Section II.  All the possible sets
of projection operators that might occur in the construction of a set of
alternative histories like (2.1) are restricted to projections on ranges of
position.  The predictions of the restricted sets agree because of
identities that express sums-over-histories in terms of operators. For
instance, 
\begin{equation}
\int_{[x^{\prime\prime}, \Delta_n, \cdots, \Delta_1, x^\prime]}
dx\, e^{iS[x(t)]/\hbar} = \left\langle x^{\prime\prime} \left|P^n_{\Delta_n}
(t_n)\cdots P^1_{\Delta_1} (t_1)\right|x^\prime\right\rangle
\label{threefour}
\end{equation}
where the sum on the left hand side is over all paths that start at
$x^\prime$ pass through the regions $(\Delta_1, \cdots, \Delta_n)$ at times
$t_1, \cdots, t_n$ and end at $x^\prime$ \cite{Cav86}.

The sum-over-histories formulation of quantum theory is not usually
discussed as a different {\it interpretation} of quantum mechanics.  But it
can be \cite{FH65, Sor97} because, like Bohm theory, it specifies a
fundamental set of variables. In effect, it posits a set selection
principle. To the extent that the quasiclassical realm in which we operate
as human observers can be described as a coarse graining of configuration
space \cite{GHup} no predictive power is lost in making this restriction.
However, the restriction is not so strong as to narrow the range of
available sets just to the quasiclassical realm.

There is some loss in convenience with a sum-over-histories formulation
because quantities like the momentum of a particle must be described in
spacetime terms --- by time of flight for example \cite{FH65}. But there is
also potential gain. A sum-over-histories restriction provides a head
start in the characterization of classicality and the explanation of its
origin (see {\it e.g.}~\cite{GH93}).  A sum-over-histories formulation of quantum
mechanics is the natural framework for investigating generalizations of
quantum mechanics that are necessary to describe spacetime alternatives
extended over time ({\it e.g.}~\cite{YT91a}) and those which
may be needed for a quantum theory of gravity \cite{Har93c,Har95c}.

\section{Is There One Interpretation of Quantum Mechanics?}

It would be interesting to investigate how many different interpretations 
of quantum
theory can be seen as restrictions of the quantum mechanics of closed
systems together with the approximations and particular descriptions of
histories that these restrictions permit.  That would be at least one way
of connecting different interpretations and a common basis for discussing
their assumptions, advantages, motivations, and limitations.

It would be equally interesting to identify interpretations of quantum
mechanics which cannot be viewed as restrictions of the quantum mechanics
of closed systems for some fundamental reason. (And not simply because they
lack the coherence to decide.) Consistent histories quantum mechanics is
logically consistent, consistent with experiment as far as is known,
consistent with textbook predictions for measurements, and applicable to the
most general physical systems. However, it may not be the only theory with these
properties. Investigations of interpretations that do not fit within its
umbrella framework may lead in different directions. 

Can we distinguish between the different interpretations that are
restrictions of the quantum mechanics of closed systems? Not by experiment
or observations.  By assumption the different interpretations
 agree on the predictions for
measurement to excellent approximations. It seems unlikely to this author
that we can settle on one interpretation by argument and discussion. 
(There is some empirical evidence for this conclusion.) There are too many
individually held opinions on the objectives to be met by the restrictions.
But neither does there seem to be a compelling need to settle among
interpretations that are restrictions of a common quantum mechanics of
closed systems

We may be able to distinguish interpretations by their utility and/or
their promise as starting points for generalizations or alternatives to
quantum theory.  For instance, Copenhagen quantum mechanics is inadequate
for cosmology.  In cosmology there is no fundamental division of the closed
system into two parts, one of which measures the other. Measurements and
observers cannot be fundamental in a theory that seeks to describe the
early universe where neither existed. In a quantum world there are
generally no variables that behave classically in all circumstances.
As another example,
sum-over-histories quantum theory may be a productive route to generalizing
usual quantum theory to incorporate the dynamical spacetime geometry of
quantum gravity \cite{Har95c}.

Many years ago, when an instructor at Princeton, I discussed my
first effort in understanding quantum mechanics \cite{Har68} with Eugene
Wigner. At the conclusion of the discussion I asked him whether I should
publish my results. Wigner explained that there were some subjects --- and
the interpretation of quantum mechanics was one of these --- that one
couldn't learn about by reading books or attending lectures. One just had
to work through them by oneself. And usually, if people took the trouble to
do this and reached a conclusion, they published a paper.  ``So'', he said,
``why shouldn't you?'' Maybe that is another reason there are so many
interpretations of quantum theory.

\section*{\bf Acknowledgments}

This work was supported in part by NSF grant PHY00-70895.

\end{document}